\documentclass[%
 aip,
 amsmath,amssymb,
reprint, twocolumn 
]{revtex4-1}

\usepackage[pdftex]{graphicx}
\usepackage[pdftex]{color}
\usepackage{dcolumn}
\usepackage{bm}

\usepackage[utf8]{inputenc}
\usepackage[T1]{fontenc}
\usepackage{mathptmx}
\usepackage{etoolbox}

\makeatletter
\def\@email#1#2{%
 \endgroup
 \patchcmd{\titleblock@produce}
  {\frontmatter@RRAPformat}
  {\frontmatter@RRAPformat{\produce@RRAP{*#1\href{mailto:#2}{#2}}}\frontmatter@RRAPformat}
  {}{}
}%
   \def\hs{\hspace}
   
   \def\le{\leqslant}
   \def\ge{\geqslant}
   \def\bs{\boldsymbol}
   \def\pder#1#2{\frac{\partial #1}{\partial #2}}

   \def\NZR#1{$#1\times10^5$ cells/mL}
   \def\MM#1{$#1$ mm}
\makeatother
\begin{document}

\preprint{AIP/123-QED}

\title[
   Formation of a Single Bioconvection Spot in {\it Euglena} Suspension Induced by Negative Phototaxis
]{
   Formation of a Single Bioconvection Spot in {\it Euglena} Suspension Induced by Negative Phototaxis
}
\author{H. Yamashita}
\email{hyamashi@hiroshima-u.ac.jp}
\affiliation{
   Graduate School of Integrated Sciences for Life, Hiroshima University, Hiroshima, Japan
}
\author{T. Yamaguchi}%
\affiliation{
   Data Science and AI Innovation Research Promotion Center, Shiga University, Shiga, Japan
}
\author{N. J. Suematsu}
\affiliation{
   Graduate School of Advanced Mathematical Sciences, Meiji University, Tokyo, Japan
}
\author{S. Izumi}
\affiliation{
   Graduate School of Integrated Sciences for Life, Hiroshima University, Hiroshima, Japan
}
\author{M. Iima}
\affiliation{
   Graduate School of Integrated Sciences for Life, Hiroshima University, Hiroshima, Japan
}

\date{\today}

\begin{abstract}
   Microorganisms are known to alter their motility in response to external stimuli.
   A typical example is the response known as taxis, which includes behaviors such as moving toward a light source (positive phototaxis) or away from it (negative phototaxis).
   In this study, we focused on bioconvection induced by the negative phototaxis of a {\it Euglena} suspension exposed to strong light from the bottom.
   Recent studies have revealed that {\it Euglena} bioconvection induced by negative phototaxis exhibits localized structure.
   In a previous study, we found that a single {\it Euglena} bioconvection spot (EBC spot) maintains its structure in a cylindrical container.
   Such a localized structure is not exclusive to {\it Euglena}, but has also reported in red tide algae, suggesting that this is a universal feature of bioconvection phenomena.
   In this paper, we investigated the formation conditions of EBC spot phenomena through experiments and numerical simulations.
   In the experiments, we examined the variation in EBC spot formation as a function of suspension height and average cell density, revealing that the critical average cell density increases with suspension height.
   In the numerical simulations, we conducted bioconvection simulations based on a model of {\it Euglena} motility that incorporated three types of behavior: the negative gravitaxis, negative phototaxis, and behavior of moving toward darker areas.
   Our model successfully reproduced the localized state of a two-dimensional EBC spot.
   Furthermore, calculations corresponding to the experimental conditions of suspension height and average cell density revealed that the critical average cell density for bioconvection decreases as the suspension height increases.
\end{abstract}

\maketitle

\section{Introduction}
   Microorganisms alter their movement in response to the surrounding environment.
   It is known that microorganisms possess a trait called taxis, including
   phototaxis\cite{hader-iseki}, where they move toward or away from light source;
   gravitaxis\cite{hader}, where they move along or against the direction of gravity;
   and chemotaxis\cite{berg}, where movement changes in response to surrounding chemical gradient.
   When microorganisms in a fluid move upward due to taxis, achieving a mass density stratification where the upper layer becomes heavier, a downward flow occurs due to Rayleigh--Taylor instability.
   When the upward movement of microorganisms and downward flow are in balance, a convective phenomenon called bioconvection occurs in the suspension\cite{wager,platt}.
   This study focuses on a type of bioconvection caused by the negative phototaxis of {\it Euglena}.

   {\it Euglena} is a unicellular organism that performs photosynthesis and exhibits spontaneous movement.
   It has a body length of approximately 50 \textmu m and swims using a single flagellum attached at the front.
   {\it Euglena} is known to exhibit various responses to light stimuli, such as negative and positive phototaxis, as well as photophobic response and photokinesis\cite{hader-iseki}.
   Phototaxis refers to movement along the light direction\cite{diehn}, with positive phototaxis meaning movement toward the light source, and negative phototaxis meaning movement away from the light source.
   It is also known that {\it Euglena} exhibits negative gravitaxis\cite{hader-hemmersbach}.

   Suematsu et al.\cite{suematsu} conducted an experiment where a suspension of {\it Euglena} ({\it E. gracilis}) was illuminated from below with strong light to induce negative phototaxis, generating bioconvection within the container.
   They discovered that bioconvection caused by negative phototaxis exhibits spatially localized bioconvection spots, suggesting that this localization requires not only vertical movement due to phototaxis, but also some form of horizontal movement.
   The term {\it bioconvection spot} describes a structural pattern analogous to a single hexagonal flow configuration characteristic of B\'{e}nard convection, distinguished by its ability to exist independently.
   The factor contributing to the horizontal movement was suggested to be {\it Euglena}'s behavior of moving toward darker areas, such as the shadows of other individuals.
   Giometto et al.\cite{giometto} conducted an experiment where they filled a linear channel with a dilute suspension of {\it Euglena} and illuminated the suspension from below using diffuse light from an LED point source, creating a spatial gradient in intensity.
   Ogawa et al.\cite{ogawa} also conducted a similar experiment using a thinner container that restricted {\it Euglena}'s movement to the horizontal plane, and adopted a spatial-gradient light environment created using a photo mask by an overhead-projector (OHP) film with a printed grayscale pattern.
   They measured the cell density or the number of {\it Euglena} at various locations within the container and demonstrated that the number of individuals was low in areas with zero and high light intensity, while a peak in the number of individuals appeared between these regions.
   Their studies revealed that {\it Euglena} has the ability to detect light gradients.
   This photoresponse is considered the main cause of the localization observed in {\it Euglena} bioconvection.

   The localized bioconvection of {\it Euglena} has been reported to occur not only in bioconvection spots, but also in a {\it single} bioconvection spot.
   Shoji et al.\cite{shoji} conducted experiments on {\it Euglena} bioconvection in an annular container as a quasi-two-dimensional system and demonstrated that a single bioconvection spot can maintain its structure even in the absence of surrounding bioconvection spots.
   In our previous study, we developed an experimental system that manipulates the local cell density of {\it Euglena} suspensions in a cylindrical container using a spatiotemporally controlled special light environment\cite{yamashita}.
   We demonstrated that a single three-dimensional bioconvection spot of {\it Euglena}, which we refer to as a {\it Euglena} bioconvection spot (EBC spot), can emerge using our system and independently maintain its structure without the surrounding cells.
   This type of localized bioconvection spot has also been reported to occur in suspensions of {\it Alexandrium fundyense}, one of the causes of red tide\cite{persson-smith}, although their potential self-aggregation behavior remains unclear.

   The localized bioconvection spot phenomenon observed in microalgae despite differences in their causes may indicate a universality of this feature.
   When the motility required for localization is clarified, it may be possible to infer the motility of microorganisms from the observed bioconvection.
   A useful approach to understanding localized bioconvection is theoretical analysis incorporating the modeling of microbial swimming and fluid dynamics.
   Various bioconvection models based on taxis, such as phototaxis\cite{vincent-hill,panda-ghorai,arrieta}, gravitaxis\cite{childress,taheri-bilgen}, gyrotaxis\cite{ghorai-hill1,ghorai-hill2,karimi-paul}, and chemotaxis\cite{lee-kim,yanaoka-nishimura}, have been studied in terms of their linear stability analysis and stable convective structures.
   However, localized convection has not been studied, and it remains unclear whether localization can emerge in these models.

   The present study aims to understand the formation mechanisms of the localized convection observed in {\it Euglena} bioconvection.
   This paper focuses on the emergence of the EBC spot, clarifies the formation conditions, and investigates its characteristics.
   We experimentally demonstrated its emergence, which depends on the suspension height and average cell density, using the setup from our previous study\cite{yamashita}.
   In addition, to theoretically investigate the localization mechanism, we developed a bioconvection model that takes into account the motility of {\it Euglena}, including negative gravitaxis, negative phototaxis, and the behavior of moving toward darker areas.
   We conducted numerical simulations of bioconvection in a simple system, a closed two-dimensional rectangular container, to demonstrate the basic characteristics of the constructed model and to investigate whether it can reproduce a two-dimensional EBC spot.

   This paper is organized as follows.
   In Section \ref{experimental-method}, we present the experimental method for observing an EBC spot using a non-uniform light environment, as well as the conditions of the {\it Euglena} suspensions observed.
   Section \ref{numerical-method} discusses the model of {\it Euglena} motility and numerical simulations of bioconvection.
   In Section \ref{experimental-results}, we show the experimental results of an EBC spot for various suspension heights and average cell densities.
   Section \ref{numerical-results} provides the characteristics of bioconvection obtained using our model and the results corresponding to the experiments.
   In Section \ref{discussion}, we discuss the implications of our findings.
   Finally, Section \ref{conclusions} concludes the paper with a summary of the main results and suggestions for future work.

\section{Methods}
   \subsection{Observations of the emergence of EBC spot varying with suspension height and average cell density\label{experimental-method}}
\begin{figure}[h]
   \centering
   \includegraphics[width=\linewidth]{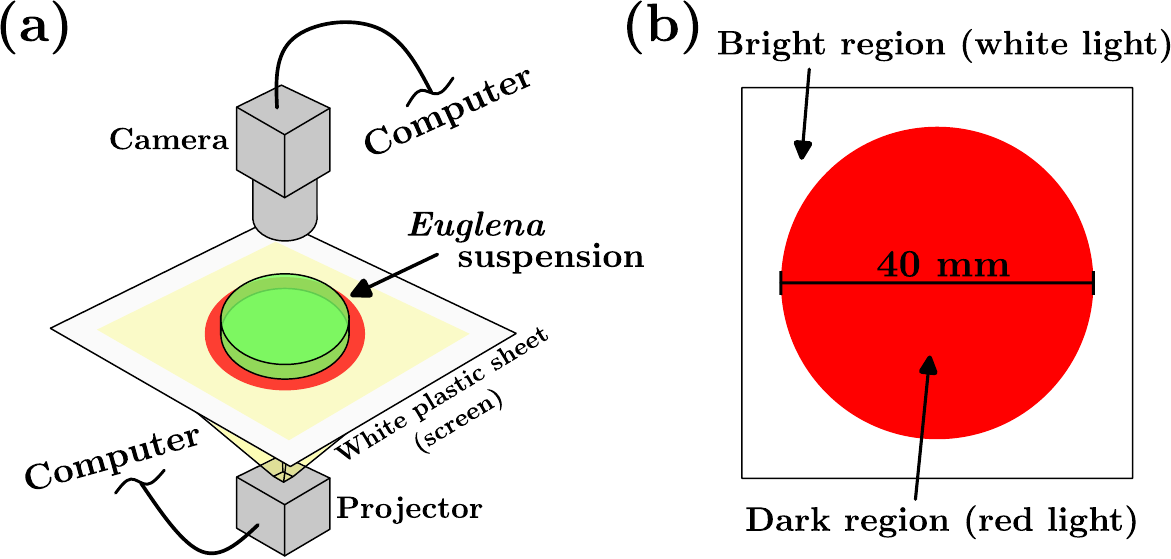}
   \caption{
      The experimental system (a) and the initial condition of a non-uniform light field (b).
   }
   \label{ExpSetup}
\end{figure}

      The experimental setup, which is the same as that in our previous study\cite{yamashita} except for the suspension height and average cell density, is as follows.
      To observe a single EBC spot, we adopted the experimental system illustrated in Fig.~\ref{ExpSetup}(a).
      {\it Euglena} suspensions in a Petri dish with a 20-mm radius, placed on a white plastic sheet, were exposed to a non-uniform light field from below.
      We controlled the light field by a computer using a projector (Pico Cube X, Felicross Co. Ltd.) as a light source.
      The change in the field was as follows.
      First, a red light (15.6 \textmu mol/m$^2$/s) was projected on a whole region of the sheet for 5 min to adapt the {\it Euglena} to the initial stimuli.
      Next, a red circled light with a 20-mm radius and a white light (55.4 \textmu mol/m$^2$/s) were projected as shown in Fig.~\ref{ExpSetup}(b).
      The radius of the red circle shrinked at a constant rate over 15 min and finally, the whole region of the sheet was illuminated by a white light.
      During the period when the red circle was shrinking and for 5 min after the circle vanished, we recorded the {\it Euglena} suspension in the container from above by a camera (CS500-C, SHODENSHA) with a macro lens (L-630, HOZAN).
      The white light used in this study was set to an intensity that induces negative phototaxis in {\it Euglena}.
      By contrast, we selected red light as the weaker light compared to the white light.
      With this setting and the above procedure, {\it Euglena} was induced to move toward the red light region.
      In a thin container (0.5-mm height) where bioconvection did not occur, it has been confirmed that when a light environment comprising a red circular region (5-mm radius) and surrounding white region is applied, the number of {\it Euglena} within the red region increases\cite{muku}.

      To investigate the characteristics of the EBC spot, we conducted experiments with suspensions of various average cell densities and heights.
      We used {\it E. gracilis} strain Z pre-cultured using Koren--Hunter medium with continuous light environment.
      They were inoculated into a 1-g/L HYPONeX aqueous solution with periodic light environment at 20$\pm$1 $^\circ$C.
      In this study, a total of twenty parameter sets were considered.
      We selected the average cell density $n_0^*=0.5$, 0.75, 1.0, 1.25, and \NZR{1.5}.
      In this paper, the dimensional quantities are denoted by a superscript $^*$.
      The suspension height was set as $L_y^*=2$, 3, 4, and \MM{5}.
      Each suspension was diluted with a HYPONeX solution, the same as that used in the inoculation.
      By measuring the absorbance, we prepared suspensions with the desired average cell density one day before the experiment.
      This was conducted to eliminate the effects of disturbances on {\it Euglena}'s conditions caused by the dilution of the suspensions.
      Just before the experiment, we measured the absorbance again to re-estimate the average cell density.
      Deviations of up to 11 \% were observed from the five selected average cell densities.
      The re-estimated values were then used as the experimental parameters.

   \subsection{Numerical simulations of {\it Euglena} bioconvection\label{numerical-method}}
   \subsubsection{A model of {\it Euglena} motility}
\begin{figure}[h]
   \centering
   \includegraphics[width=\linewidth]{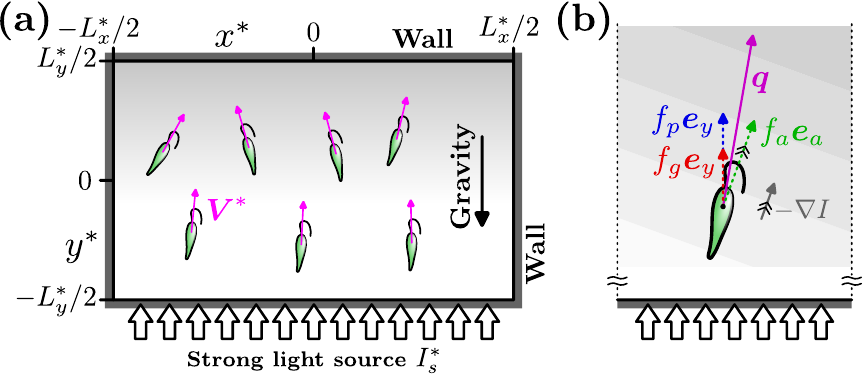}
   \caption{
      The flow configuration (a) and a schematic of the motility model (b).
      The grayscale gradient represents the light-intensity field.
   }
   \label{Model}
\end{figure}

      We assumed that the {\it Euglena} suspension is illuminated by strong light from below, as shown in Fig.~\ref{Model}(a).
      The model of the flux $\bs{J}^*$ of local cell density $n^*$ is represented as
      \begin{equation}
         \bs{J}^* = n^*\bs{u}^*-D^*\nabla^*n^*+n^*\bs{V}^*
         .
      \end{equation}
      Here, $\bs{u}^*$ and $D^*$ represent fluid velocity and diffusion coefficient of {\it Euglena} cells, respectively.
      $\bs{V}^*$ is average swimming velocity of the local population of {\it Euglena} due to the motility, defined as $\bs{V}^*=V_d^*\bs{e}_q$.
      $V_d^*$ represents the swimming speed of {\it Euglena}.
      In this study, the swimming direction vector $\bs{e}_q$ is defined as the unit vector of $\bs{q}$, which is determined by the vector sum of the following motilities: the negative gravitaxis, negative phototaxis, and behavior of moving toward darker areas (see Fig.~\ref{Model}(b)).
      Thus,
      \begin{equation}
         \bs{e}_q = \frac{\bs{q}}{|\bs{q}|}
         ,
      \end{equation}
      \begin{equation}
         \bs{q} = (f_g+f_p)\bs{e}_y+f_a\bs{e}_a
         .
         \label{qvec}
      \end{equation}
      Here, $\bs{e}_y$ is the unit vector pointing vertically upward.
      The $x^*$ axis is oriented horizontally, and the $y^*$ axis vertically (Fig.~\ref{Model}(a)).
      The first term on the right-hand side of Eq. (\ref{qvec}) represents negative gravitaxis, second term represents negative phototaxis, and third term represents the behavior of moving toward darker areas.
      $f_g$, $f_p$, and $f_a$ are the susceptibilities of motilities, defined as
      \begin{equation}
         f_g = c_g
         ,\hs{5pt}
         f_p = c_pI
         ,\hs{5pt}
         f_a = c_aI
         ,
      \end{equation}
      where $c_g$, $c_p$, and $c_a$ are constants.
      The properties of the motility model are determined by the ratio of these constants.
      $I$ represents the non-dimensionalized light-intensity field based on the intensity of the light source $I_s^*$.
      The dimensional light intensity $I^*$ was evaluated by the Lambert--Beer's law in the case of illumination to the suspension from below:
      \begin{equation}
         I^*(t^*,x^*,y^*) = I_s^*\exp\left[ -\alpha^*\int_{-L_y^*/2}^{y^*} n^*(t^*,x^*,y') dy' \right]
         \label{lightfield}
      \end{equation}
      where $\alpha^*$, $n_0^*$, and $L_y^*$ represent the absorption coefficient, average cell density, and suspension height, respectively.
      The light source is placed at the bottom ($y^*=-L_y^*/2$; Fig.~\ref{Model}(a)).
      $\bs{e}_a$ appearing in the third term governing the behavior of moving toward darker areas in Eq. (\ref{qvec}) was defined as 
      \begin{equation}
         \bs{e}_a = -\frac{\nabla I}{|\nabla I|}
      \end{equation}
      using the gradient of the light-intensity field.
      Thus, in this model, {\it Euglena} cells swim with velocity $\bs{V}^*$ in the weighted direction determined by the relative strengths of each type of motility.

\begin{figure}[h]
   \centering
   \includegraphics[width=\linewidth]{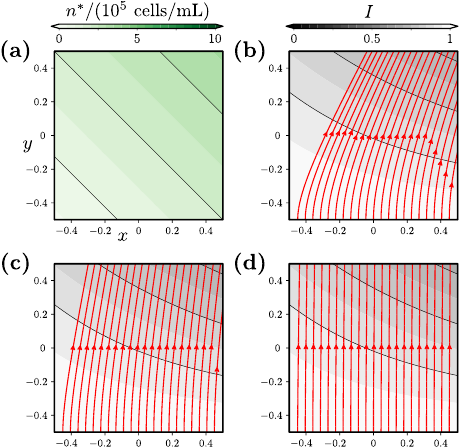}
   \caption{
      The base cell density distribution (a) and the trajectories along the swimming direction vector $\bs{e}_q$, shown as red lines with arrows, (b--d).
      The light-intensity fields (colormaps rendered in grayscale) in (b--d) are identical.
      (a) $c_g:c_p:c_a=0.05:0.05:1$, (b) $c_g:c_p:c_a=1:1:1$, and (c) $c_g:c_p:c_a=1:1:0.05$.
   }
   \label{SwimDire}
\end{figure}

      Figure \ref{SwimDire} illustrates how the net swimming direction $\bs{e}_q$ is determined by the ratio of the motility constants ($c_g:c_p:c_a$).
      In the case where the cell density increases linearly from the bottom left to the top right of the container (Fig.~\ref{SwimDire}(a)), $n^*=n_0^*[1+0.8(x^*+y^*)/L_y^*]$, the light-intensity field calculated using Eq. (\ref{lightfield}) results in a grayscale colormap, as shown in Fig.~\ref{SwimDire}(b--d).
      The red lines with arrows in Fig.~\ref{SwimDire}(b--d) represent the trajectories along $\bs{e}_q$, corresponding to cases where the ratios of the motility constants are (a) $c_g:c_p:c_a=0.05:0.05:1$, (b) $c_g:c_p:c_a=1:1:1$, and (c) $c_g:c_p:c_a=1:1:0.05$.
      When the contribution of $c_a$, associated with the gradient of the light-intensity field, is high, our model exhibits movement toward the darker region (the top-right corner of the container in Fig.~\ref{SwimDire}(a)).
      The movement becomes almost entirely upward as the contribution of $c_a$ decreases.
      We note that, when the cell density distribution is uniform, the movement becomes purely upward regardless of the values of the motility constants, even if only $c_a$ has a non-zero value.
      This is because in this case, $I$ depends only on $y^*$, and the gradient of $I$ aligns with $\bs{e}_y$.

   \subsubsection{Governing equations and boundary conditions}
      Assuming that the {\it Euglena} suspension is an incompressible fluid, the continuity equation is expressed as
      \begin{equation}
         \nabla^*\cdot\bs{u}^* = 0
      \end{equation}
      The fluid motion follows the Navier--Stokes equations under the Boussinesq approximation:
      \begin{equation}
         \pder{\bs{u}^*}{t^*} + (\bs{u}^*\cdot\nabla^*)\bs{u}^* = -\frac{\nabla^* p^*}{\rho^*}+\frac{\mu^*}{\rho^*}\nabla^{*2}\bs{u}^*-n^*\frac{\delta_\rho^*}{\rho^*}v^*g^*\bs{e}_y
         \label{dimpsi}
      \end{equation}
      Here, $p^*$, $\rho^*$, and $\mu^*$ represent the pressure field, fluid density, and fluid viscosity, respectively.
      $\delta_\rho^*$ represents the density difference between the fluid ($\rho^*$) and {\it Euglena} ($\rho_E^*$), defined as $\delta_\rho^*=\rho_E^*-\rho^*$.
      $v^*$ is the volume of {\it Euglena} and $g^*$ is the gravitational acceleration.
      The conservation equation of cell density is expressed by
      \begin{equation}
         \pder{n^*}{t^*} = -\nabla^*\cdot\bs{J}^* = -(\bs{u}^*\cdot\nabla^*)n^*+D^*\nabla^{*2}n^*-\nabla\cdot(n^*\bs{V}^*)
         \label{dimcel}
         .
      \end{equation}

      We consider the bioconvection in a two-dimensional rectangular container with the width $L_x^*$ and height $L_y^*$, as shown in Fig.~\ref{Model}(a).
      Equations (\ref{dimpsi}) and (\ref{dimcel}) are non-dimensionalized using $t^*=tL_y^{*2}/D^*$, $\bs{x}^*=\bs{x}L_y^*$, $\bs{u}^*=\bs{u}D^*/L_y^*$, $p^*=p\mu^*D^*/L_y^{*2}$, $n^*=nn_0^*$.
      In addition, the stream function $\psi$ is adopted: $\bs{u}\cdot\bs{e}_x=\partial\psi/\partial y$, $\bs{u}\cdot\bs{e}_y=-\partial\psi/\partial x$.
      Thus, the governing equations are
      \begin{multline} 
         \pder{}{t}(\nabla^2\psi)+\left( \pder{\psi}{y}\pder{}{x}-\pder{\psi}{x}\pder{}{y} \right)\nabla^2\psi \\
         = {\rm Sc}\left[ \nabla^2(\nabla^2\psi)+{\rm Ra}\pder{n}{x} \right]
         ,
         \label{nondimpsi}
      \end{multline} 
      \begin{equation}
         \pder{n}{t}  = -\left( \pder{\psi}{y}\pder{}{x}-\pder{\psi}{x}\pder{}{y} \right)n+\nabla^2n-\nabla\cdot[ n{\rm Pe}\bs{e}_q(I) ]
         ,
         \label{nondimcel}
      \end{equation}
      \begin{equation}
         I(t,x,y) = \exp\left[ -\alpha\int_{-1/2}^{y} n(t,x,y') dy' \right]
         .
         \label{nondimlight}
      \end{equation}
      The dimensionless parameters appear in the governing equations:
      \begin{equation}
         {\rm Sc} = \frac{\mu^*}{\rho^*D^*}
         ,\hs{5pt}
         {\rm Ra} = \frac{n_0^*L_y^{*3}\delta_\rho^*v^*g^*}{\mu^*D^*}
         ,\hs{5pt}
         {\rm Pe} = \frac{V_d^*L_y^*}{D^*}
         ,\hs{5pt}
         \alpha = \alpha^*n_0^*L_y^*
         .
      \end{equation}
      Table.~\ref{properties} shows the properties for determining the dimensionless parameters.
      The volume of {\it Euglena} was assumed to be that of a cylinder with a radius of 5 \textmu m and a height of 50 \textmu m.
      The absorbance coefficient was adopted from the values used to estimate the average cell density in the experiments.
      ${\rm Sc}$ is fixed at $7.7$.
      ${\rm Ra}$, ${\rm Pe}$, and $\alpha$ are determined by $n_0^*$ and $L_y^*$.
      The parameter sets selected in the experiments are within the ranges of $6.0\lessapprox{\rm Ra}\lessapprox290$, $1.5\lessapprox{\rm Pe}\lessapprox3.8$, and $0.028\lessapprox\alpha\lessapprox0.21$, respectively.

      The boundary conditions are no-slip conditions for fluid flow and no cell flux at the walls of the rectangular container.
      Therefore,
      \begin{equation}
         \psi = \pder{\psi}{x} = 0
         ,\hs{5pt}
         \pder{n}{x} = n{\rm Pe}(\bs{e}_q\cdot\bs{e}_x)
         \hs{5pt}{\rm at}\hs{5pt}
         x = \pm \frac{A}{2}
         ,
         \label{bcx}
      \end{equation}
      \begin{equation}
         \psi = \pder{\psi}{y} = 0
         ,\hs{5pt}
         \pder{n}{y} = n{\rm Pe}(\bs{e}_q\cdot\bs{e}_y)
         \hs{5pt}{\rm at}\hs{5pt}
         y = \pm \frac{1}{2}
         .
         \label{bcy}
      \end{equation}
      $A$ $(=L_x^*/L_y^*)$ represents the aspect ratio of the container.

\begin{table}[h]
   \centering
   \caption{Properties of the {\it Euglena} suspension.}
   \label{properties}
   \begin{ruledtabular}
   \begin{tabular}{rl}
      Cell diffusion coefficient $D^*$ & $1.3\times10^{-7}$ m$^2$/s [Ref.~\onlinecite{giometto}] \\
      Average swimming speed $V_d^*$ & $10^{-4}$ m/s [Ref.~\onlinecite{rodrigues}] \\
      Volume of an individual $v^*$ & $4\times10^{-15}$ m$^3$ \\ 
      Absorption coefficient $\alpha^*$ & $2.8\times10^{-10}$ m$^2$/cells \\
      Density of a {\it Euglena} cell $\rho_E^*$ & $1.05\times10^3$ kg/m$^3$ [Ref.~\onlinecite{lebert}] \\
      Fluid density $\rho^*$ & $10^3$ kg/m$^3$ \\
      Fluid viscosity $\mu^*$ & $10^{-3}$ Pa$\cdot$s \\
      Gravitational acceleration $g^*$ & $9.8$ m/s$^2$ \\
   \end{tabular}
   \end{ruledtabular}
\end{table}

   \subsubsection{Numerical method}
      To solve the governing equations (\ref{nondimpsi})--(\ref{nondimlight}) under the boundary conditions (\ref{bcx}) and (\ref{bcy}), we adopted the spectral collocation method for spatial discretization and the Crank--Nicolson method for time discretization.
      The stream function and cell density were expanded based on Chebyshev polynomials:
      \begin{equation}
         \psi = \sum_{k=0}^{M_x}\sum_{l=0}^{M_y}\psi_{kl}\widetilde{T}_k\left( \frac{2}{A}x \right)\widetilde{T}_l\left( 2y \right)
         ,
         \label{specpsi}
      \end{equation}
      \begin{equation}
         n = \sum_{k=0}^{M_x}\sum_{l=0}^{M_y}n_{kl}T_k\left( \frac{2}{A}x \right)T_l\left( 2y \right)
         ,
         \label{speccel}
      \end{equation}
      where $M_x$ and $M_y$ represent the expansion orders in the $x$- and $y$-directions, respectively.
      We used the Gauss--Lobatto points as collocation points.
      $T_i(\xi)$ represents Chebyshev polynomials, and $\widetilde{T}_i(\xi)$ represents modified Chebyshev polynomials that satisfy the boundary conditions of $\psi$ given by (\ref{bcx}) and (\ref{bcy}): $\widetilde{T}_i(\xi)=T_i(\xi)(1-\xi^2)^2$.

      In Section \ref{numerical-results}, we present calculations for aspect ratios $A=2$, 8, 10, 13.3, and 20.
      In the case of $A=2$, the steady-state solutions were obtained as follows.
      The discretized governing equations can be symbolically expressed as:
      \begin{equation}
         \frac{d}{dt}\bs{\phi} = \bs{f}(\bs{\phi})
         .
         \label{dphidt}
      \end{equation}
      Here, $\bs{\phi}$ represents a vector comprising the coefficients $\psi_{kl}$ and $n_{kl}$ appearing in Eqs. (\ref{specpsi}) and (\ref{speccel}):
      \begin{equation}
         \bs{\phi}=(\psi_{00},\psi_{01},\cdots,\psi_{M_xM_y},n_{00},n_{01},\cdots,n_{M_xM_y})
         .
      \end{equation}
      The steady-state solution $\overline{\bs{\phi}}$ satisfies $\bs{f}(\overline{\bs{\phi}})=\bs{0}$.
      By applying the Newton--Raphson method to Eq. (\ref{dphidt}), $\overline{\bs{\phi}}$ can be obtained.
      In addition, a linear stability analysis was conducted.
      By introducing a perturbation $\bs{\varepsilon}$ to $\overline{\bs{\phi}}$ and substituting $\bs{\phi}'=\overline{\bs{\phi}}+\bs{\varepsilon}$ into Eq. (\ref{dphidt}), the linearized equation is obtained:
      \begin{equation}
         \frac{d}{dt}\bs{\varepsilon} = \mathsf{G}\bs{\varepsilon}
         ,\hs{5pt}
         \mathsf{G} \equiv \left[ \pder{\bs{f}}{\bs{\phi}} \right]_{\bs{\phi}=\overline{\bs{\phi}}}
         .
      \end{equation}
      The eigenvalues of the Jacobian matrix $\mathsf{G}$ were calculated, and the maximum real part of these eigenvalues ($s_{\rm max}$), i.e., the maximum linear growth rate, was evaluated.
      The sign of $s_{\rm max}$ determines the stability.

      For cases other than $A=2$, time-evolution calculations were performed with two types of initial distributions of the cell density $n$, where $\psi(t=0)$ was set to zero.
      One distribution involves placing a high-cell-density region in the upper center of the computational domain, approximately replicating the cell distribution when the darker region vanished in the experiments manipulating the local cell density:
      \begin{equation}
         \widehat{n} = \exp\left[ -\left( \frac{x}{\sqrt{2}\sigma} \right)^2-\left( \frac{y-1/2}{\sqrt{2}\sigma} \right)^2 \right]+\frac{1}{\kappa}
         .
      \end{equation}
      We set $\sigma=0.2$ and $\kappa=250$ in the calculations.
      The initial condition was set to
      \begin{equation}
         n(t=0) = \widehat{n}'
         ,\hs{5pt}
         \widehat{n}' \equiv \frac{\gamma\widehat{n}V}{\int_V\gamma\widehat{n}dV}
         ,
         \label{spotbasedinit}
      \end{equation}
      since the initial conditions should satisfy $\int_V ndV=V$.
      $\gamma(x,y)$ ($0.999\le\gamma\le1.001$) represents a small position-dependent disturbance and $V$ is the computational domain.
      The other is based on a trivial steady-state solution $\overline{n}$:
      \begin{equation}
         n(t=0) = \overline{n}'
         ,\hs{5pt}
         \overline{n}' \equiv \frac{\gamma\overline{n}V}{\int_V\gamma\overline{n}dV}
         ,
         \label{stesolbasedinit}
      \end{equation}
      where
      \begin{equation}
         \overline{n} = \frac{{\rm Pe}\exp({\rm Pe}y)}{2\sinh({\rm Pe}/2)}
         .
         \label{stesol}
      \end{equation}
      We performed time-evolution calculations for at least 90 min in dimensional form to compare them with the experimental observations.

      For the verification of numerical simulations, we fixed the parameters as $c_g:c_p:c_a=0.05:0.05:1,A=1$, \NZR{n_0^*=3}, and \MM{L_y^*=5}, and compared the maximum value of $|\psi|$ in the computational domain ($|\psi|_{\rm max}$) for various $M_y$ ($=M_x$).
      Assuming that sufficient resolution of calculations is provided in the case of $M_y=40$, and using the obtained $|\psi|_{\rm max}$ as a reference, the differences were 0.44 \% for $M_y=10$, 0.018 \% for $M_y=20$, and 0.0021 \% for $M_y=24$.
      Considering the balance between computation time and accuracy, we adopted $M_y=20$ for the subsequent calculations and set $M_x=AM_y$ according to the aspect ratio.

\begin{figure}[h]
   \centering
   \includegraphics[width=0.7\linewidth]{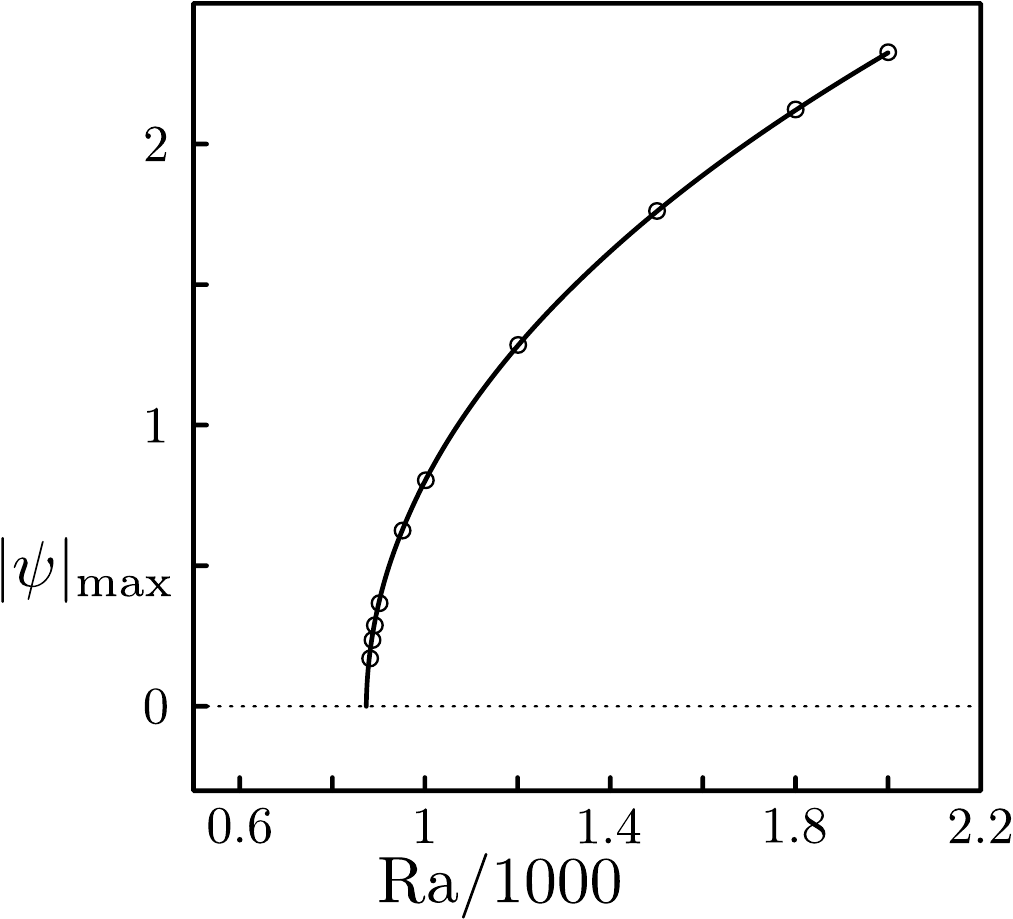}
   \caption{
      Calculations of bioconvection considering only gravitaxis.
      The parameters were set to $c_g:c_p:c_a=1:0:0$, $A=2$, ${\rm Sc}=1$, and ${\rm Pe}=1$.
      Circles: Taheri and Bilgen\cite{taheri-bilgen} and solid line: the present study.
   }
   \label{Validation}
\end{figure}

      For validation, we compared our results with the numerical simulations of gravitactic bioconvection conducted by Taheri and Bilgen\cite{taheri-bilgen}.
      In our model, the setting of $c_g:c_p:c_a=1:0:0$ is equivalent to their model.
      The $|\psi|_{\rm max}$ obtained for $A=2,{\rm Sc}=1$, and ${\rm Pe}=1$ were plotted against ${\rm Ra}$ on the horizontal axis in Fig.~\ref{Validation}.
      It shows that our results (solid line) align with the values they obtained (circles).

\section{Results}
   \subsection{Experimental results\label{experimental-results}}
\begin{figure}[h]
   \centering
   \includegraphics[width=\linewidth]{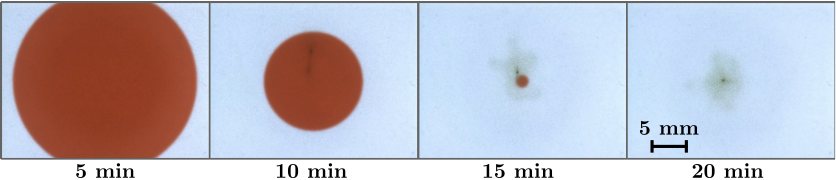}
   \caption{
      The representative observation using our experimental system.
      These figures show the progression over time from left to right, with snapshots at 5, 10, 15, and 20 min, respectively.
      \MM{L_y^*=5} and \NZR{n_0^*=0.99}.
   }
   \label{ReRe}
\end{figure}

      Figure \ref{ReRe} shows a representative result of EBC spot formation through the manipulation of local cell density using the non-uniform light environment in Sec.~\ref{experimental-method}.
      The suspension height and average cell density were \MM{L_y^*=5} and \NZR{n_0^*\approx1.0}, respectively.
      The progression of time is shown from left to right in the figure, with snapshots at 5, 10, 15, and 20 min from the start of observation.
      In our experimental method, {\it Euglena} in the suspension are guided to the red light region, avoiding the white light, and as this region shrinks, a high-cell-density region forms in the center of the container.
      During the contraction of the red region, a high-cell-density region (shown as a less saturated region) appears within the red circle, as seen in the snapshot at 15 min.
      Our method allowed bioconvection to occur in part of the container, and a single EBC spot was achieved as shown in the figure.
      The emerged EBC spot maintained its structure even after the red region vanished.

\begin{figure}[h]
   \centering
   \includegraphics[width=\linewidth]{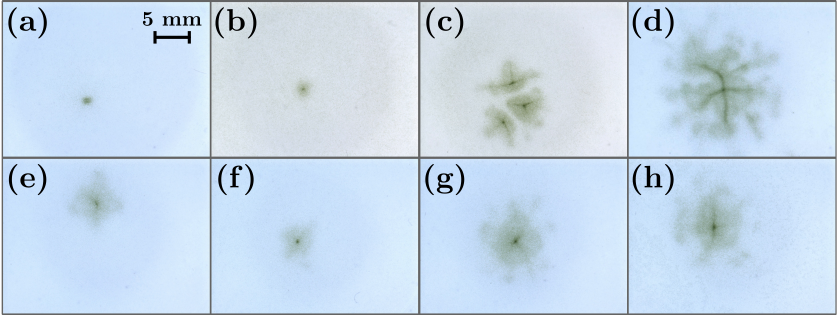}
   \caption{
      The EBC spots emerged at various $L_y^*$ and $n_0^*$ values.
      (a--d) \NZR{n_0^*\approx1.5}: (a) \MM{L_y^*=2}, (b) \MM{3}, (c) \MM{4}, and (d) \MM{5}.
      (e--h) \MM{L_y^*=5}: (e) \NZR{n_0^*=0.47}, (f) \NZR{0.74}, (g) \NZR{1.01}, and (h) \NZR{1.25}.
   }
   \label{Dep}
\end{figure}

\begin{figure}[h]
   \centering
   \includegraphics[width=\linewidth]{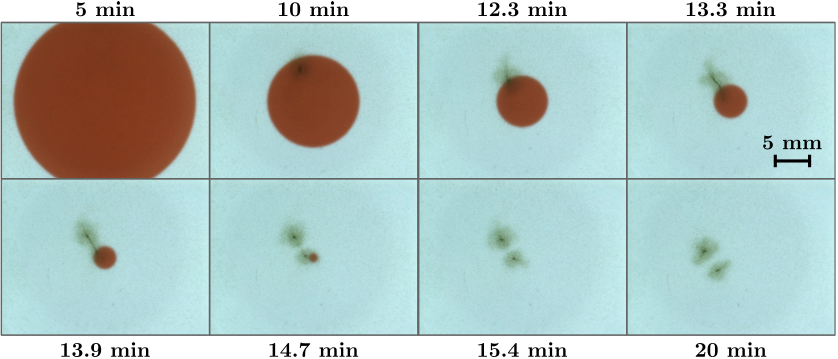}
   \caption{
      The split into two EBC spots.
      \MM{L_y^*=3} and \NZR{n_0^*=1.57}.
   }
   \label{Split}
\end{figure}

      For a total of twenty parameter sets, observational experiments of an EBC spot were conducted, varying the suspension height and average cell density.
      Figures \ref{Dep}(a--d) shows the differences in an EBC spot due to changes in suspension height, with the average cell density fixed at \NZR{n_0^*\approx1.5}.
      The snapshot of the EBC spot that appeared at the height of \MM{2} is shown in Fig.~\ref{Dep}(a), and the suspension height increases sequentially from (a) to (d).
      These figures indicate that the EBC spot increases in size and splits depending on the height.
      The details of the split observed at \NZR{1.57} and \MM{L_y^*=3} are shown in Fig.~\ref{Split}.
      It indicates that an EBC spot had already formed during the process of the red circular region shrinking.
      Subsequently, the EBC spot extended across the boundary of the red region, and after the red region vanished, two EBC spots were eventually observed.
      At the height of \MM{5} shown in Fig.~\ref{Dep}(d), the structure extends over a wide region of the container.
      Figures \ref{Dep}(e--h) and (d) show EBC spots observed with a fixed height of \MM{5}, with the average cell density increasing sequentially.
      Although not as distinct as in the case of height variation, a similar feature was observed.
      This indicates that the size of the EBC spot increases in the order of Figs. \ref{Dep}(e), (g), and (h).
      The spot sizes in Figs. \ref{Dep}(e) and (f) are approximately the same as are those in Figs. \ref{Dep}(g) and (h).


\begin{figure}[h]
   \centering
   \includegraphics[width=0.7\linewidth]{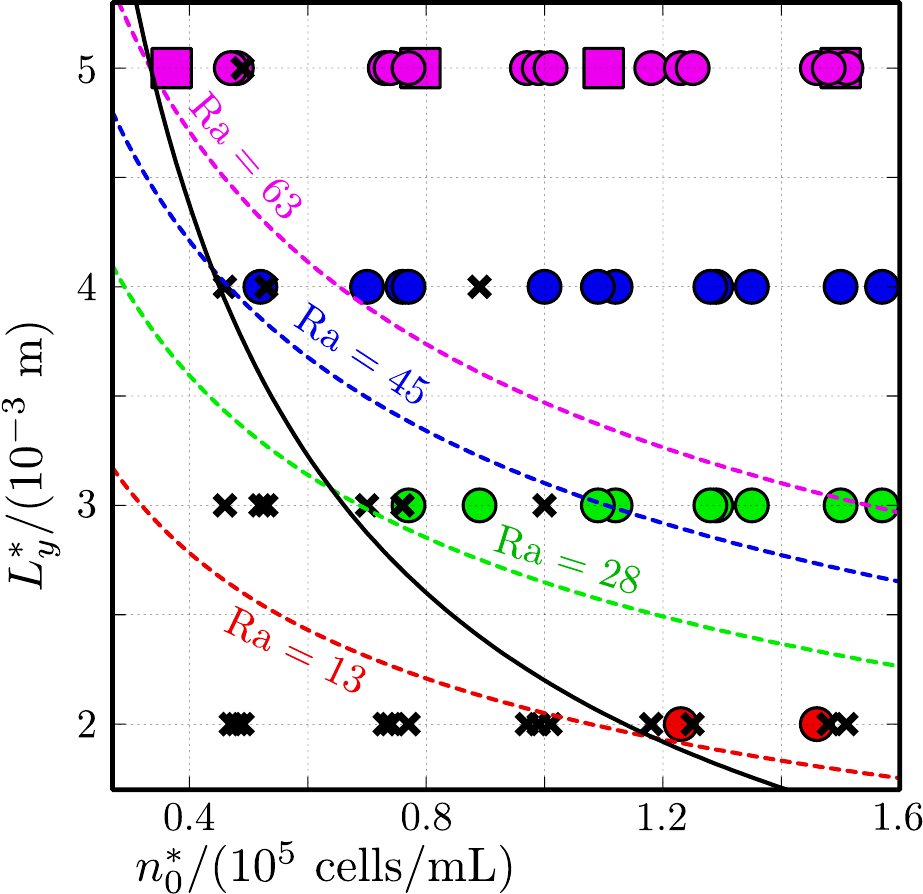}
   \caption{
      The emergence of EBC spots depending on $L_y^*$ and $n_0^*$.
      Circles and cross marks indicate whether the EBC spots emerged.
      Squares: Yamashita et al.\cite{yamashita}
      Dotted curves: the contours of ${\rm Ra}$.
      Solid curve: $L_y^*=2.2n_0^{*-0.75}$.
   }
   \label{NzrHeight}
\end{figure}

      In Fig. \ref{NzrHeight}, we summarize whether EBC spots emerged for each of the twenty parameter sets selected in the experiments.
      A circle mark indicates cases where bioconvection was observed, and a cross mark indicates cases where it was not.
      A square mark represents our previous results\cite{yamashita}.
      A total of sixty points, more than the number of parameter sets, are plotted in the figure, because three observations were conducted for each parameter set.
      Dashed curves represent the contours of ${\rm Ra}$.
      Our experimental results demonstrated that the critical ${\rm Ra}$ varies with each height of the suspension, and it tends to increase as the height becomes larger.
      A solid curve representing $L_y^*\propto n_0^{*-0.75}$ was plotted to pass through the critical average cell density for each height of the {\it Euglena} suspension.
      Considering that the contour line of ${\rm Ra}$ is represented by $L_y^*\propto n_0^{*-1/3}$ and that of constant cell count is represented by $L_y^*\propto n_0^{*-1}$, we found that the solid curve is relatively close to the constant-cell-count contour.

   \subsection{Numerical results\label{numerical-results}}
\begin{figure}[h]
   \centering
   \includegraphics[width=\linewidth]{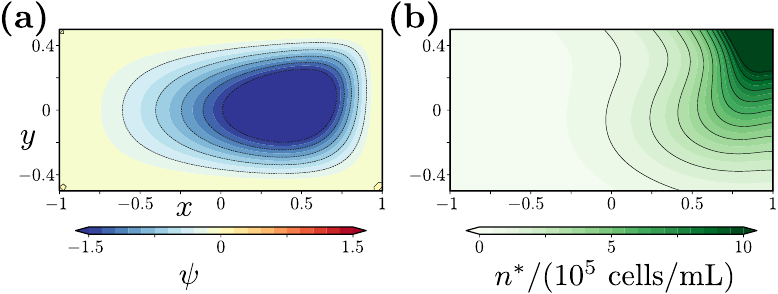}
   \caption{
      The single bioconvection roll appeared in the container with $A=2$.
      \MM{L_y^*=5} and \NZR{n_0^*=2}.
      $c_g:c_p:c_a=0.05:0.05:1$.
      The distributions of stream function $\psi$ (a) and dimensional cell density (b).
   }
   \label{CgCpCa-FP-PsiCel}
\end{figure}

\begin{figure}[h]
   \centering
   \includegraphics[width=\linewidth]{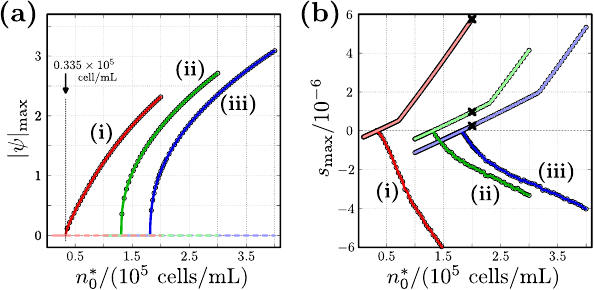}
   \caption{
      The emergence of bioconvection for $A=2$ using our model in the cases of several parameter set.
      (a) The maximum absolute value of the stream function $\psi$.
      Solid lines represent stable solutions, while dotted lines represent unstable solutions.
      (b) The maximum linear growth rate $s_{\rm max}$ for trivial (light red, green, and blue) and non-trivial (dark colors) solutions plotted in (a).
      (i) $c_g:c_p:c_a=0.05:0.05:1$, (ii) $c_g:c_p:c_a=1:1:1$, and (iii) $c_g:c_p:c_a=1:1:0.05$.
   }
   \label{CgCpCa-FP}
\end{figure}

      To understand the characteristics of our bioconvection model, numerical simulations were performed for \MM{L_x^*=10} and \MM{L_y^*=5}.
      Focusing on the vicinity of the critical average cell density where a nontrivial solution first emerges from a trivial one, we selected three ratios of the motility constants: (i) $c_g:c_p:c_a=0.05:0.05:1$, where the behavior of moving toward darker areas is dominant; (ii) $c_g:c_p:c_a=1:1:1$, where three motilities equally contribute; and (iii) $c_g:c_p:c_a=1:1:0.05$, where upward movement is dominant.
      In all three ratios, the flow field that appears at the lowest $n_0^*$ is a flow with a single convection roll in the container, as shown in Fig.~\ref{CgCpCa-FP-PsiCel}.
      This feature, where the single-roll flow first emerges, has been reported by Taheri and Bilgen\cite{taheri-bilgen} in a bioconvection model that considers only the upward swimming of microorganisms.
      The occurrence of the single-roll flow for each ratio was calculated, and the results were plotted in Fig.~\ref{CgCpCa-FP}(a), with $n_0^*$ on the horizontal axis and $|\psi|_{\rm max}$ on the vertical axis.
      Figure \ref{CgCpCa-FP}(b) shows the maximum linear growth rate, $s_{\rm max}$, corresponding to trivial and non-trivial solutions plotted in Fig.~\ref{CgCpCa-FP}(a).
      This indicates that $s_{\rm max}$ of the trivial solutions (represented by light red, green, and blue lines) transitions from negative to positive as $n_0^*$ increases.
      The non-trivial solutions (represented by dark colors) emerge when $s_{\rm max}$ of the trivial solutions exceeds zero.
      Additionally, the corresponding $s_{\rm max}$ are negative, indicating that these solutions, corresponding to the single-roll flow shown in Fig.~\ref{CgCpCa-FP-PsiCel}, are stable.
      Regardless of the ratio of the motility constants, it was found that the single-roll flow field undergoes a pitchfork bifurcation, leading the onset of bioconvection once the critical average cell density is exceeded.
      The values of the critical average cell density are \NZR{0.335} (${\rm Ra}=63.2$) for the ratio (i), \NZR{1.31} (${\rm Ra}=247$) for the ratio (ii), and \NZR{1.82} (${\rm Ra}=343$) for the ratio (iii), respectively.
      We found that as the contribution of the behavior of moving toward darker areas increases, the critical average cell density decreases.


\begin{figure}[h]
   \centering
   \includegraphics[width=\linewidth]{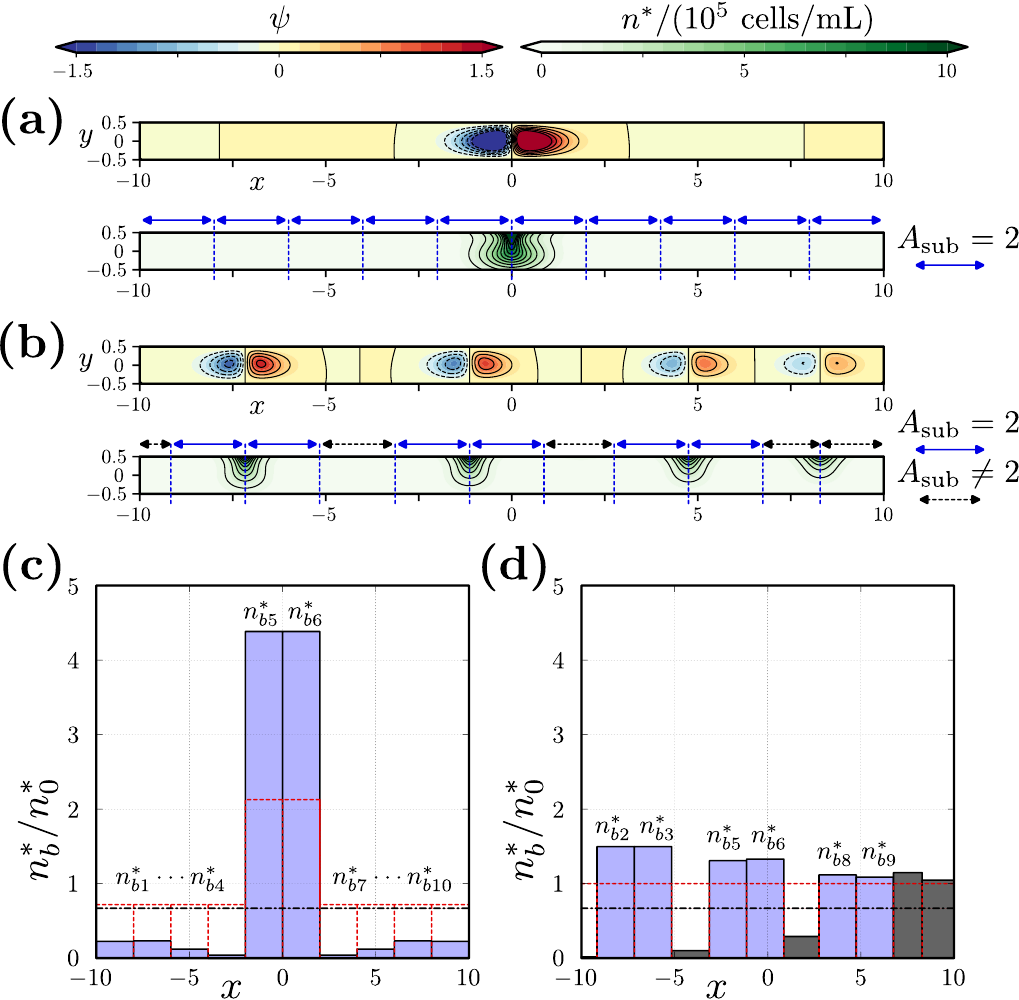}
   \caption{
      A single 2D-EBC spot (a) and several 2D-EBC spots (b) emerged at the dimensional time of about 220 min.
      $A=20$, \MM{L_y^*=5}, and \NZR{n_0^*=0.5}.
      $c_g:c_p:c_a=0.05:0.05:1$.
      The initial conditions were set to $\widehat{n}'$ for (a) and to $\overline{n}'$ for (b).
      (c) and (d) represent the histograms of $n_b^*$ within the range indicated by the arrow (blue solid: $A_{\rm sub}=2$ and black dashed: $A_{\rm sub}\ne2$) shown in (a) and (b).
      (c) and (d) correspond to (a) and (b), respectively.
      The initial conditions of $n_b^*$ are plotted with red dashed boxes, while the states at 220 min are shown as filled boxes (blue: $A_{\rm sub}=2$ and gray: $A_{\rm sub}\ne2$).
      Black dash-dotted lines correspond to \NZR{n_0^*=0.335}, which is the critical average cell density for $A=2$.
   }
   \label{LBC-TE}
\end{figure}

      To investigate whether our model can reproduce a two-dimensional EBC spot (2D-EBC spot) comprising a pair of convection rolls, numerical simulations were performed in a horizontally wide container.
      The ratio of motility constants was set to $c_g:c_p:c_a=0.05:0.05:1$, where the behavior of moving toward darker areas is dominant, making bioconvection more likely to occur.
      A container with a height of \MM{L_y^*=5} and width of \MM{L_x^*=100} was selected ($A=20$).
      The average cell density was set to \NZR{n_0^*=0.5}, corresponding to the value used in the observational experiments.
      Figure \ref{LBC-TE}(a) shows the calculation results for the initial condition (\ref{spotbasedinit}), which replicate the experiment, while (b) shows the results for the initial condition (\ref{stesolbasedinit}), representing the trivial solution.
      Time-evolution calculations over a dimensional time of approximately 220 min were performed for each initial condition.
      They demonstrate that a 2D-EBC spot appears and maintains its structure at the center of the container in case (a), while multiple 2D-EBC spots appear within the container in case (b).
      When the average cell density was slightly lower, at \NZR{0.3}, similar calculations showed that a 2D-EBC spot appears, as seen in Fig.~\ref{LBC-TE}(a), under the initial condition replicating the experiments (the condition (\ref{spotbasedinit})).
      By contrast, no convection occurs in the container when starting with the trivial solution (the condition (\ref{stesolbasedinit})).

      To understand the characteristics of the two distributions of cell density shown in Figs.~\ref{LBC-TE}(a) and (b), the container was divided into subdomains such that each region contained a single bioconvection roll, and the average cell density within each subdomain, $n_b^*$, was calculated.
      In Fig. \ref{LBC-TE}(a), the 2D-EBC spot is located at the center of the container, so the container was evenly divided into ten subdomains with an aspect ratio of $A_{\rm sub}=2$ (from left to right, $n_b^*$ was labeled as $n_{b1}^*,n_{b1}^*,\cdots,n_{b10}^*$).
      In Fig. \ref{LBC-TE}(b), boundary lines were drawn to divide each 2D-EBC spot into subdomains.
      Except for the rightmost 2D-EBC spot, subdomains with $A_{\rm sub}=2$ were assigned, while the remaining regions were set to fill the gaps.
      This resulted in the container being divided to 11 subdomains in total (from left to right, $n_b^*$ was labeled as $n_{b1}^*,n_{b1}^*,\cdots,n_{b11}^*$).
      $n_b^*$ is presented as histograms in Figs.~\ref{LBC-TE}(c) and (d).
      The vertical axis shows $n_b^*$ normalized by the overall average cell density of the container, $n_0^*$.
      The open boxes with red dashed lines represent the initial conditions, while the filled boxes correspond to the estimated results based on the cell density distributions in Figs.~\ref{LBC-TE}(a) and (b).
      The dash-dotted lines indicate the critical average cell density of \NZR{0.335} for $A=2$, as shown in Fig.~\ref{CgCpCa-FP}.
      From Fig.~\ref{LBC-TE}(c), which corresponds to the result for the initial condition with a high cell density region placed at the center of the container, it was found that after 220 min of time evolution, the average cell densities $n_{b5}^*$ and $n_{b6}^*$ within the central subdomains increased to approximately 4.4 times $n_0^*$.
      This indicates that approximately 90 \% of the organisms were concentrated in the central subdomains of the container.
      The cell densities $n_{b1}^*,\cdots,n_{b4}^*,n_{b7}^*,\cdots,n_{b9}^*$ and $n_{b10}^*$ in the subdomains outside the center decreased significantly, falling well below the critical average cell density for $A=2$.
      This behavior was also observed for the initial condition corresponding to the trivial solution.
      As seen in Fig.~\ref{LBC-TE}(d), the cell densities $n_{b2}^*,n_{b3}^*,n_{b5}^*,n_{b6}^*,n_{b8}^*$ and $n_{b9}^*$ in the subdomains where 2D-EBC spots formed exceeded the dash-dotted line.

\begin{figure}[h]
   \centering
   \includegraphics[width=0.7\linewidth]{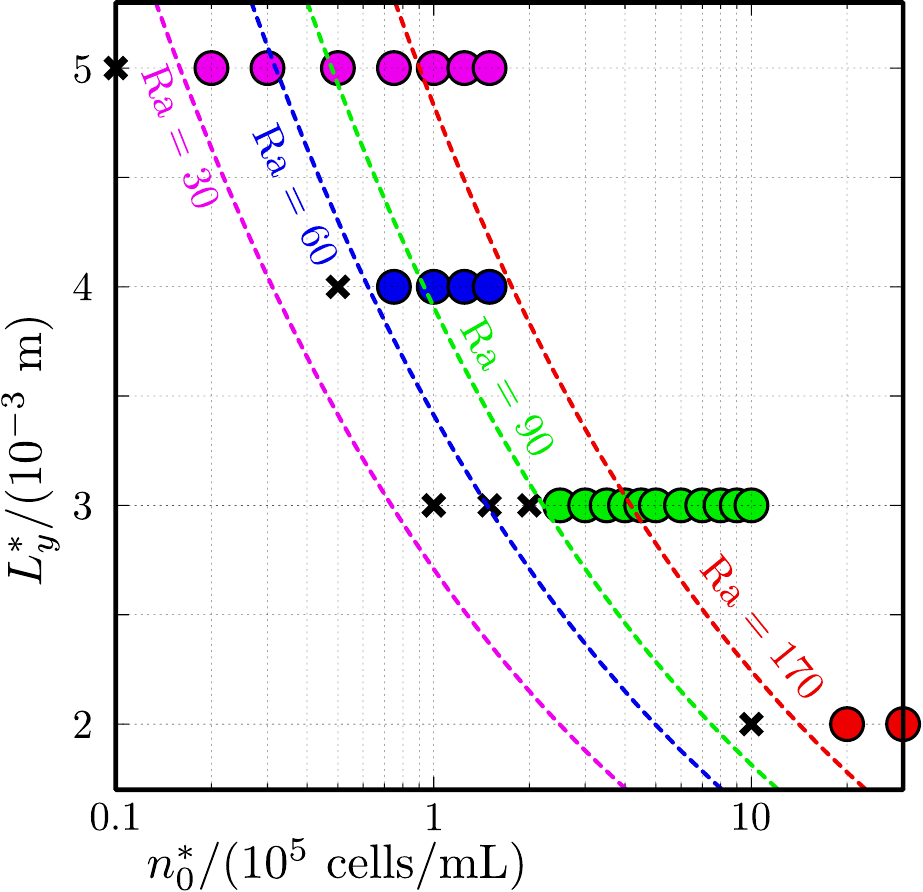}
   \caption{
      The emergence of 2D-EBC spots in our model depending on $L_y^*$ and $n_0^*$.
   }
   \label{VsExp}
\end{figure}

\begin{figure}[h]
   \centering
   \includegraphics[width=\linewidth]{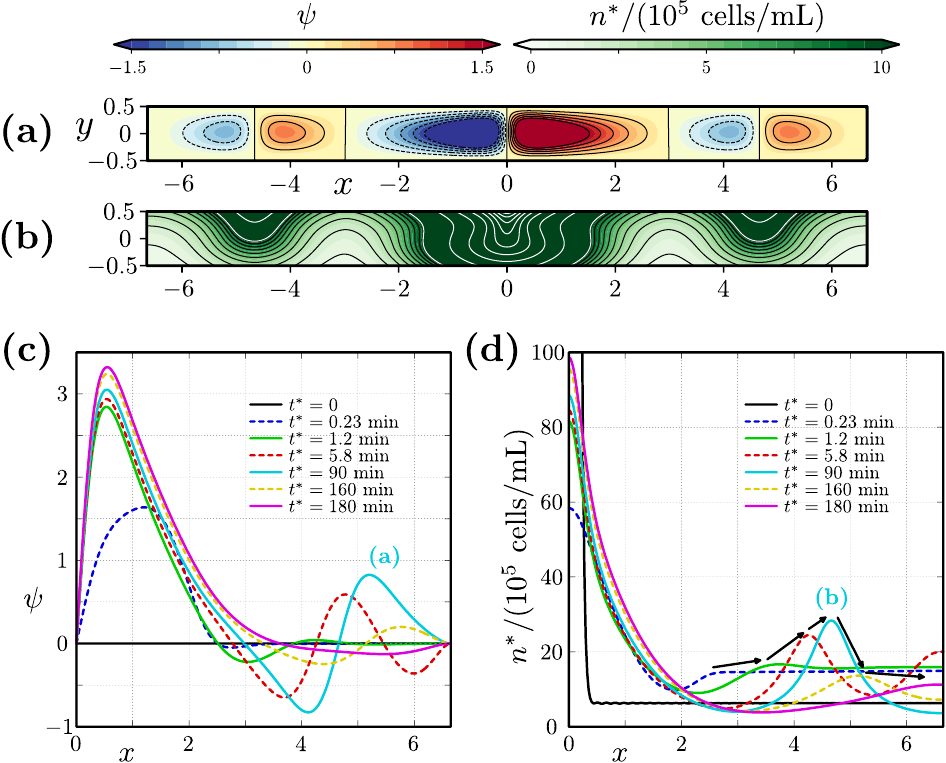}
   \caption{
      Several 2D-EBC spots formed in a container for $A=40/3$, \MM{L_y^*=3}, and \NZR{n_0^*=10} at $t^*=90$ min.
      (a) The distribution of the stream function.
      (b) The distribution of the dimensional cell density.
      (c) The time evolution of the horizontal profile ($x\ge0$) of the stream function along $y=0$.
      (d) The time evolution of the dimensional cell density at the top of the container ($y=1/2$).
   }
   \label{SimSeveralSpots}
\end{figure}

      Numerical calculations of our bioconvection model were performed for the suspension height, container width, and average cell density corresponding to the experimental results in Fig.~\ref{NzrHeight}.
      Based on the correspondence with the observation time in the experiment, we conducted time-evolution calculations in dimensional form for 90 min under the initial condition that replicate the experiment (the condition (\ref{spotbasedinit})), and the presence or absence of 2D-EBC spots at that time was confirmed.
      The results are summarized in Fig.~\ref{VsExp}.
      The horizontal axis is shown on a logarithmic scale and dashed curves represent the contour of ${\rm Ra}$.
      This figure indicates that as the suspension height increases, the critical ${\rm Ra}$ tends to decrease.

      Figure \ref{SimSeveralSpots} illustrates the flow field and cell density distribution observed in numerical simulations of 2D-EBC spot splitting under the condition of \MM{L_y^*=3} and \NZR{n_0^*=10}.
      Time-evolution calculations were performed starting from the initial condition (\ref{spotbasedinit}).
      The distributions of the stream function and cell density at 90 min (dimensional time) are shown in Figs.~\ref{SimSeveralSpots}(a) and (b), respectively.
      Three 2D-EBC spots are observed within the container.
      The time evolution of the horizontal profile of the stream function along $y=0$ and the cell density along $y=1/2$ are shown in Figs.~\ref{SimSeveralSpots}(c) and (d), respectively.
      Due to the symmetry of the distributions, only the region $x\ge0$ is plotted.
      A 2D-EBC spot originating from the initial high-density region appeared at the center of the container, followed by the formation of new 2D-EBC spots away from the initial one.
      These newly formed 2D-EBC spots gradually moved toward the side walls and disappeared over a period extending beyond the observation time of 90 min in experiments, up to 180 min (see the arrows in Fig.~\ref{SimSeveralSpots}(d)).

\section{Discussion\label{discussion}}
   By controlling the local cell density within the container using our experimental system, we succeeded in accumulating the cell density near the container center.
   It led to the emergence of bioconvection (EBC spot) near the center, as shown in Figs.~\ref{ReRe} and \ref{Dep} in Sec.~\ref{experimental-results}.
   Depending on the average cell density, the structure could be maintained.
   This structural maintenance observed in the experiments can be interpreted based on the horizontal movement suggested by Suematsu et al.\cite{suematsu}, specifically, the behavior of moving toward darker areas.
   Once a high cell density region is formed, the light intensity within this region decreases due to the shadows of {\it Euglena} individuals compared to the surrounding dilute regions.
   After being carried by downward flows due to the Rayleigh--Taylor instability, the individuals are not simply diffused, but instead return to the center of the EBC spot (the darker region).
   As a result, the convective structure can be maintained.

   As shown in Figs.~\ref{Dep}(c) and \ref{Split}, a single EBC spot was not always observed for all parameter sets in this experimental system.
   For a suspension height of \MM{4}, an EBC spot was observed at $n_0^*=0.5$--\NZR{1.0}, but at high average cell densities ($1.25$ and \NZR{1.5}), several EBC spots were observed after the red region shrank.
   Suematsu et al.\cite{suematsu} demonstrated that the mean distance between adjacent EBC spots does not depend on the average cell density for \MM{L_y^*=2} and \NZR{n_0^*\gtrapprox5}.
   Taking this into account, it can be inferred that there is an upper limit on the size (or the number of cells) of the EBC spot for each suspension height, and the splitting occurred at higher average cell densities for $3$ and \MM{4}, because the size (or the number of cells) exceeded the limit.
   A qualitative comparison between experimental and numerical results (Figs.~\ref{Split} and \ref{SimSeveralSpots}) revealed agreement in the observation that at higher average cell densities, a single EBC spot transitions to multiple cells.
   However, the splitting behavior differed between experiments and numerical simulations.
   In experimental results, a single EBC spot appeared to split into two, whereas in the simulations, new EBC spots emerged on both sides of the existing cell.
   Furthermore, the simulations indicated that the split state is transient.
   This implies that the state in which multiple EBC spots appear in experiments is not stationary and may eventually merge into a single EBC spot.

   From the experimental results (Fig.~\ref{NzrHeight}), we found that the critical ${\rm Ra}$ increases with the height of the suspension.
   It should be noted that not only ${\rm Ra}$, but also ${\rm Pe}$ changes with height.
   As suggested by the form of the steady solution (\ref{stesol}) (top heaviness is enhanced as ${\rm Pe}$ increases) and the finding of Taheri and Bilgen\cite{taheri-bilgen}, an increase in the ${\rm Pe}$ promotes destabilization.
   In other words, the critical ${\rm Ra}$ tends to decrease as the system height increases.
   The reason why the experimental results do not follow this trend is unclear.
   The curve of the critical average cell density (solid curve in Fig.~\ref{NzrHeight}) inferred from the observations appears to closely follow the contour of constant cell count.
   This observation suggests a biologically driven mechanism: bioconvection forms in a region when a specific number of individuals is localized within that area.
   In our model, all properties of {\it Euglena} suspensions are treated as averaged quantities, and the ratio of motility constants is fixed.
   It may be necessary to consider biological characteristics, such as variations in the motility contributions, average swimming speed, and diffusion coefficient, based on local cell density and position.
   Regarding diffusion coefficient, Childress et al.\cite{childress} have pointed out that local cell density, position, and anisotropy likely influence it in reality.
   There may also exist motility traits of {\it Euglena} that are unaccounted for or yet to be identified within the framework of the present model.
   For instance, if {\it Euglena} possesses self-aggregation properties induced by attractant substances, as observed in {\it Tetrahymena}\cite{kohidai}, it could more effectively enhance local high-density aggregation.

   In the bioconvection model of {\it Euglena} adopted in the present study, one notable feature is that the critical ${\rm Ra}$ for the onset of bioconvection decreases when the behavior of moving toward darker areas is dominant, as shown in Fig.~\ref{CgCpCa-FP}.
   The movement toward darker areas includes horizontal motility, which refers to motion directed toward regions of high cell density (see Fig.~\ref{SwimDire}).
   As the result, compared to cases involving only vertical upward movement, the initially fluctuating trivial solution is expected to become more strongly destabilized.
   In other words, the maximum linear growth rate is higher when the behavior of moving toward darker areas is incorporated, compared to that in cases where it is absent.
   At an average cell density of \NZR{2.0}, where bioconvection occurred in all three ratios of motility constants (i--iii), case (iii) showed the smallest $s_{\rm max}$ (see cross marks in Fig.~\ref{CgCpCa-FP}(b)).
   The values for case (ii) was approximately four times larger than that of case (iii), while case (i) was approximately 20 times larger.

   As shown in Fig.~\ref{LBC-TE}, our bioconvection model was found to reproduce localized bioconvection states under parameters similar to those used in experiments.
   The localized state was characterized by approximately 10 \% of the organisms being located in non-convective regions (comprising 80 \% of the container's domain) and a single 2D-EBC spot, containing 90 \% of the organisms, forming in the remaining 20 \% of the container's domain.
   The average cell densities in the non-convective regions were lower than the critical average cell density for $A=2$, making it reasonable to assume that bioconvection could not occur in those subdomains.
   Since the boundary conditions on the sides of each subdomain are not wall boundary conditions, a more precise comparison is required for a similar discussion in a system with periodic boundary conditions applied in the horizontal direction.

   The time-evolution calculations over a finite period for \MM{L_y^*=5} and \NZR{n_0^*=0.3} suggest that bistability may be observed in our bioconvection model.
   Bistability in bioconvection has also been reported in experiments\cite{shoji,yamashita} and in another bioconvection model\cite{taheri-bilgen}.
   Based on previous studies, the bistability observed in our model appears reasonable.
   However, to prove its existence, stability analysis is indispensable.
   In this study, due to computational cost constraints, we did not perform calculations of steady-state solutions and linear stability analysis for systems with large aspect ratios.
   In future research, we plan to conduct linear stability analysis of this model for wide containers.
   This approach will enable us to verify the existence of bistability observed within a finite observation period using the bioconvection model.

   We found that there are other differences in behavior of bioconvection between experiments and numerical simulations.
   For instance, microbial cluster ejection from an EBC spot, as observed in our previous study\cite{yamashita} and in Figs.~\ref{ReRe} and \ref{Dep}(c--g), was not confirmed in the numerical calculations.
   Additionally, the traveling of an EBC spot within the container, as shown by Shoji et al.\cite{shoji}, was not observed.
   Once localized convection occurred in our model, it generally remained stationary.
   As shown in Fig.~\ref{SimSeveralSpots}, there are cases where newly formed 2D-EBC spots move and eventually disappear.
   The present model is relatively simple; for instance, it does not yet incorporate hydrodynamic interactions between cells.
   Such interactions within microbial populations may induce instabilities and potentially contribute to the discrepancies observed between experimental and numerical results.
   Lauga has analytically demonstrated that an aligned state of microorganisms in suspension is inherently unstable, regardless of whether the microorganisms are pusher-type or puller-type\cite{lauga}.
   Ishikawa et al. reported the hydrodynamic interactions between cells lead to weak aggregation: pusher-type cells in close proximity tend to weakly align in opposite directions, while puller-type cells tend to align in the same direction\cite{ishikawa}.
   Oyama et al. reported that a population of puller-type swimmers in a confined container with top and bottom walls forms a traveling wave of cell density in the vertical direction\cite{oyama}.
   By incorporating these interactions into the motility model of microorganisms as changes in average swimming velocity and diffusion coefficient based on local cell density, it is expected that a more explanatory {\it Euglena} bioconvection model consistent with the observations can be developed.

\section{Conclusions\label{conclusions}}
   This study focuses on the emergence of {\it Euglena} bioconvection spots as an extension of our previous research.
   Through observational experiments, we investigated the dependence of EBC spot formation on the suspension height and average cell density.
   Derived from the experiments, the relationship between the critical average cell density and suspension height showed that the critical average cell density increases with the suspension height.
   Additionally, we developed a mathematical model that incorporates three types of motility observed in {\it Euglena}: negative gravitaxis, negative phototaxis, and the behavior of moving toward darker areas.
   Based on this model, we performed numerical simulations of bioconvection.
   Our model successfully reproduced the localized state of a single EBC spot, as observed in experiments.
   However, the phenomenon of cluster ejection from an EBC spot was not numerically confirmed.
   Using parameters corresponding to experiments, we conducted numerical calculations to determine the conditions under which bioconvection occurs.
   The results revealed that the critical average cell density decreases with increasing suspension height.
   While both the experiments and the model provided insights into localized bioconvection of {\it Euglena}, some aspects showed agreement while others did not.
   This discrepancy may be attributed to motility factors that are not yet considered in the model.
   Future studies will focus on these aspects.

   This paper represents a first step toward bridging fluid dynamics and biology.
   It is evident that further biological studies to clarify the motility characteristics of {\it Euglena} will enable improvements to the model.
   Conversely, we anticipate that model-based research can explore various motility patterns, observe the resulting convection structure, and identify the specific motility traits necessary for cluster ejection phenomena.

\begin{acknowledgments}
   This work was partially supported by JSPS KAKENHI Grant Number 21H05311.
   We would like to thank Editage (www.editage.jp) for English language editing.
\end{acknowledgments}

\section*{Data Availability Statement}
   The data that support the findings of this study are available from the corresponding author upon reasonable request.


\end{document}